\documentclass[twocolumn,showpacs,preprintnumbers,amsmath,amssymb,showkeys]{revtex4-1}
\usepackage{bm}
\usepackage{amssymb,amsmath}
\usepackage[dvips]{graphicx}

\begin{document}

\title{Anomalous elasticity in a disordered layered XY model}

\author{Fawaz Hrahsheh}

\author{Thomas Vojta}
\affiliation{Department of Physics, Missouri University of Science and Technology, Rolla, MO 65409, USA}

\begin{abstract}
We investigate the effects of layered quenched disorder on the behavior of planar magnets, superfluids,
and superconductors by performing large-scale Monte-Carlo simulations of a three-dimensional
randomly layered XY model. Our data provide numerical evidence for the recently predicted anomalously elastic
(sliding) intermediate phase between the conventional high-temperature and low-temperature phases.
In this intermediate phase, the spin-wave stiffness perpendicular to the layers vanishes in the thermodynamic
limit  while the stiffness parallel to the layers as well as the spontaneous magnetization are nonzero. In addition,
the susceptibility displays unconventional finite-size scaling properties. We compare
our Monte-Carlo results with the theoretical predictions, and we discuss possible experiments
in ultracold atomic gases, layered superconductors and in nanostructures.
\end{abstract}

\date{\today}
\pacs{67.85.Hj,74.40.-n,75.10.Nr}

\maketitle

\section{Introduction}

Extended defects can be found in a wide variety of condensed matter systems.
For example, realistic materials often contain one-dimensional and two-dimensional
defects in the form of dislocation lines or grain boundaries.
Recent progress in
nano-technology also allows researchers to custom-design artificial structures with
analogous properties.
Extended defects in systems of ultracold atomic gases can  be created by means of
one-dimensional or two-dimensional disordered optical lattices.

Extended defects are larger than the usual finite-size impurities and are thus
harder to ``average out.'' Consequently, they  have a much greater influence
on the thermodynamic behavior of the system in question. This was first established
on the example of the  McCoy-Wu model, a two-dimensional disordered classical
Ising model whose disorder is perfectly correlated in one of the two dimensions,
i.e., it takes the form of parallel line defects. McCoy and Wu \cite{McCoyWu68,McCoyWu68a,McCoyWu69,McCoy69}
demonstrated in a series of papers that this model exhibits an exotic phase
transition at which the magnetic susceptibility is infinite over an entire
temperature range while the specific heat is smooth.
Fisher \cite{Fisher92,Fisher95} later used a strong-disorder renormalization group to show
that the critical point is of infinite-randomness type and is accompanied by strong
(power-law) Griffiths singularities \cite{Griffiths69,ThillHuse95,GuoBhattHuse96,RiegerYoung96}.
Ising models with plane defects, i.e., with perfect disorder correlations in \emph{two}
rather than one dimensions display even stronger
disorder effects: instead of being sharp, the phase transition is smeared over a range of
temperatures \cite{Vojta03b,SknepnekVojta04}.

The effective dimensionality of the defects forms the basis of a classification
\cite{VojtaSchmalian05,Vojta06} of phase transitions in systems with quenched disorder.
Three classes can be distinguished: (i) If the defect dimensionality is
below the lower critical dimension $d_c^-$ of the problem, the disordered system has
a conventional critical point with exponentially weak Griffiths singularities.
(ii) If the defect dimensionality is exactly equal to the lower critical
dimension, the critical point of the disordered system is of infinite-randomness
type and accompanied by strong power-law  Griffiths singularities.
(iii) If the defects are above the lower critical dimension,
individual regions can order independently, leading to a smearing of the global
phase transition.

The case of two-dimensional (planar) defects in systems with XY symmetry is
of particular conceptual and experimental importance. Theoretically,
XY systems with perfect disorder correlations in two dimensions are right at
the boundary between cases (ii) and (iii) in the above classification. True long-range
order on individual two-dimensional ``slabs'' is forbidden by the Mermin-Wagner theorem
\cite{MerminWagner66}; however, these regions undergo a Kosterlitz-Thouless transition
to a quasi long-range ordered phase \cite{KosterlitzThouless73}. Experimentally, order
parameters with XY symmetry occur not only in planar magnets but also in superconductors
and superfluids. The fate of the XY phase transition with two-dimensional defects is
thus of great interest for magnetic and superconducting multilayers as well as
ultracold atomic gases in one-dimensional disordered optical lattices.

Recently, two simultaneous publications \cite{MGNTV10,PekkerRefaelDemler10} investigated this question
theoretically. They predicted that the conventional high-temperature and low-temperature phases
of a randomly layered XY system are separated by an anomalously elastic intermediate phase.
In this exotic ``sliding'' phase which is part of the Griffiths region, the spin-wave
(or superfluid) stiffness parallel to the
layers is nonzero while the stiffness perpendicular to the layers vanishes.

In this paper, we report the results of large-scale Monte-Carlo simulations of a
three-dimensional randomly layered XY model which provide support for
the phase transition scenario predicted in Refs.\ \cite{MGNTV10,PekkerRefaelDemler10}.
In particular, we give numerical evidence for
existence of the anomalously elastic intermediate phase. Our paper in organized
as follows. In Sec.\ \ref{sec:model}, we define the randomly layered XY model.
We briefly summarize the theoretical predictions in Sec.\ \ref{sec:theory}. Section
\ref{sec:simulations} is devoted to the Monte-Carlo simulations and their results.
We conclude in Sec.\ \ref{sec:conclusions}.

\section{Randomly layered XY model}
\label{sec:model}

In the following, we formulate the problem in the language of the XY ferromagnet.
The results will apply to all phase transitions having O(2) or U(1) order parameters, if expressed
in terms of the appropriate variables.

\begin{figure}
\includegraphics[width=8cm]{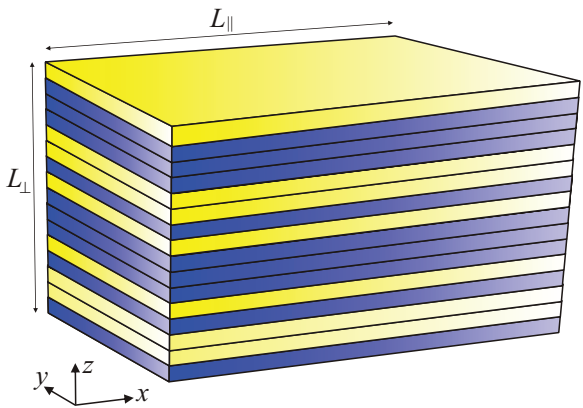}
\caption{(Color online) Sketch of the layered XY model (\ref{Eq:Hamiltonian}). Layers of two different materials
(represented by different interaction strengths $J^\parallel$) are arranged in a random sequence.}
\label{Fig:layeredmagnet}
\end{figure}

We consider a three-dimensional magnet consisting
of a random sequence of layers made up of two different materials as shown in Fig.\
\ref{Fig:layeredmagnet}.
Its Hamiltonian is a three-dimensional classical XY model defined on a lattice of
perpendicular size $L_\perp$ (in the $z$ direction) and in-plane size $L_\parallel$
(in the $x$ and $y$ directions). It reads
\begin{equation}
H = - \sum_{\mathbf{r}} J^{\parallel}_z \, (\mathbf{S}_{\mathbf{r}} \cdot \mathbf{S}_{\mathbf{r}+\hat{\mathbf{x}}}
                                        +\mathbf{S}_{\mathbf{r}} \cdot \mathbf{S}_{\mathbf{r}+\hat{\mathbf{y}}} )
    - \sum_{\mathbf{r}} J^{\perp}_z \, \mathbf{S}_{\mathbf{r}} \cdot\mathbf{S}_{\mathbf{r}+\hat{\mathbf{z}}}
    .
\label{Eq:Hamiltonian}
\end{equation}
Here, $\mathbf{S}_{\mathbf{r}}$ is a two-component unit vector on lattice site
$\mathbf{r}$, and  $\hat{\mathbf{x}}$, $\hat{\mathbf{y}}$, and $\hat{\mathbf{z}}$ are
the unit vectors in the coordinate directions. The interactions within
the layers, $J^{\parallel}_z$, and between the layers, $J^{\perp}_z$, are both positive
and independent random functions of the perpendicular coordinate $z$.
For simplicity, we take all $J^{\perp}_z$ to be identical,
$J^{\perp}_z \equiv J^{\perp}$ \footnote{The full disordered phase transition scenario emerges
even if only one of the two interactions, $J^{\parallel}_z$ and $J^{\perp}_z$, is random because
the other interaction picks up randomness under renormalization.},
while the $J^{\parallel}_z$ are drawn from a binary probability distribution
\begin{equation}
P(J^{\parallel})=(1-c)\, \delta(J^{\parallel} - J_u) + c\, \delta(J^{\parallel} - J_l)
\label{BinaryDist}
\end{equation}
with $J_u > J_l$. Here, $c$ is the concentration of the ``weak'' layers while $1-c$ is the concentration of the ``strong'' layers.

\begin{figure}
\includegraphics[angle=0,width=8cm,clip]{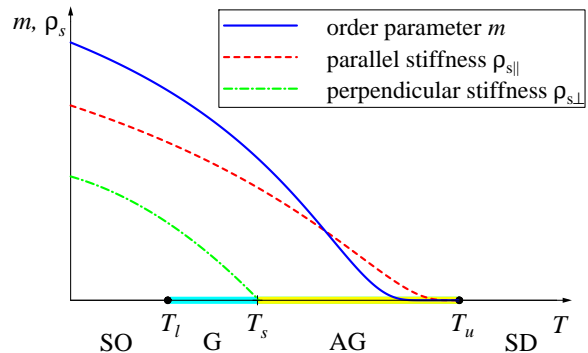}
\caption{(Color online) Schematic behavior of the order parameter (spontaneous magnetization) $m$ and the stiffnesses
$\rho_{s,\parallel}$ and $\rho_{s,\perp}$ vs. temperature $T$ for any bounded disorder
distribution. SD and SO denote the conventional strongly disordered and ordered phases,
respectively. The Griffiths region (bounded by $T_u$ and $T_l$) consists of the
``non-anomalous'' (G) and the anomalous (AG) Griffiths phases. For an unbounded
distribution, $T_u\rightarrow\infty$.}
\label{Fig:theory}
\end{figure}

Let us discuss the thermodynamics of the randomly layered XY model (\ref{Eq:Hamiltonian})
qualitatively. If the temperature $T$ is above the upper Griffiths temperature $T_u$ (defined as
the critical temperature of a hypothetical clean system having $J^{\parallel}_z \equiv J_u$ for all $z$),
the model is in a conventional paramagnetic phase (denoted ``strongly disordered'' in the phase diagram
shown in Fig.\ \ref{Fig:theory}).
Analogously, for temperatures below the lower Griffiths temperature $T_l$ (defined as
the critical temperature of a system having $J^{\parallel}_z \equiv J_l$ for all $z$),
the model is in a conventional ferromagnetic (``strongly ordered'') phase. The most interesting
temperature range is the Griffiths phase between $T_u$ and $T_l$. Here,
rare thick slabs (rare regions) of strong ($J_z^\parallel = J_u$) layers can show
local ferromagnetism while the bulk is still nonmagnetic. Individual such slabs
cannot be truly long-range ordered \cite{MerminWagner66} but they can develop quasi
long-range order via individual Kosterlitz-Thouless phase transitions \cite{KosterlitzThouless73}.
The exotic behavior predicted in Refs.\ \cite{MGNTV10,PekkerRefaelDemler10} arises from
the interplay between the quenched disorder and the Kosterlitz-Thouless
physics of these strongly interacting slabs.

\section{Optimal fluctuation theory}
\label{sec:theory}

In this section, we summarize the results of the optimal fluctuation theory of Ref.\ \cite{MGNTV10}.
A slab (rare region) consisting of $L_{RR}$ consecutive strong layers exists with an exponentially
small probability
$w(L_{RR}) \sim (1-c)^{L_{RR}} = \exp(-\tilde c L_{RR})$ with $\tilde c = -\ln(1-c)$.
It undergoes a Kosterlitz-Thouless phase transition at a temperature $T_{KT}(L_{RR})$
which can be estimated from finite-size scaling via $T_u - T_{KT} \sim L_{RR}^{-1/\nu}$.
(Here, $\nu \approx 0.6717$ \cite{CHPV06} is the correlation-length
critical exponent of a clean
3D planar (XY) magnet.)
Consequently, at a given temperature $T$, all rare regions of thickness $L_{RR} < L_c(T) \sim (T_u-T)^{-\nu}$
are (locally) in the paramagnetic phase  while those having thicknesses $L_{RR} > L_c(T)$
are in the quasi-long-range ordered phase.

We first consider the behavior of a single rare region (slab) in the quasi-long-range ordered phase.
According to Kosterlitz-Thouless theory \cite{KosterlitzThouless73}, its spin correlation function
$C(\mathbf{r})$ decays as a non-universal power of
the distance $|\mathbf{r}|$,
\begin{equation}
C(\mathbf{r}) \sim |\mathbf{r}|^{-\eta} \qquad (|\mathbf{r}| \to \infty)~.
\label{eq:KT-power}
\end{equation}
The exponent $\eta$ takes the value 1/4 right at the Kosterlitz-Thouless transition and
behaves as $1/L_{RR}$ in the limit of $L_{RR} \to \infty$. Its thickness dependence
can thus be modeled as $\eta(L_{RR}) = \frac 1 4 L_c(T)/L_{RR}$.
The power-law correlations (\ref{eq:KT-power}) lead to a
nonlinear magnetization-field curve
\begin{equation}
m \sim H^{\eta/(4-\eta)}~, \label{eq:m-H-KT}
\end{equation}
which implies that the magnetic susceptibility of a single slab
in the quasi-long-range ordered phase is infinite.

We now turn to thermodynamic observables of the full, randomly layered system
in the Griffiths phase $T_u>T>T_l$.
The spin-wave stiffness $\rho_s$ is defined via the work required to twist the spins
on two opposite boundaries of the sample by a relative angle $\theta$. In the limit of
small $\theta$ and large system size, the free-energy density $f$ depends
on $\theta$ as
\begin{equation}
 f(\theta)-f(0)=\frac{1}{2}\rho_s\left({\theta}/{L}\right)^2
 \label{eq:stiff}
\end{equation}
which defines $\rho_s$.
As our randomly layered XY model is anisotropic, we need to distinguish the
parallel spin-wave stiffness $\rho_s^\parallel$ from the perpendicular spin-wave
stiffness $\rho_s^\perp$. To determine $\rho_s^\parallel$, we apply twisted boundary
conditions at $x=0$ and $x=L_\parallel$ and set $L=L_\parallel$ in (\ref{eq:stiff}) while
the boundary conditions are applied at $z=0$ and $z=L_\perp$ (using $L=L_\perp$) to find
$\rho_s^\perp$.

When the twist is applied in $x$-direction, all layers in the sample have same
boundary conditions. The total free energy cost due to the twist is thus simply
given by a sum over all layers. As only slabs that are in the quasi-long-range
ordered phase have a nonzero stiffness, the total parallel stiffness reads
\begin{equation}
\rho_{s,\parallel} \sim \int_{L_c(T)}^\infty dL_{RR} \, w(L_{RR})\,
\rho_{s,RR}(L_{RR})~.
\label{eq:rho_s_parallel-integral}
\end{equation}
Here, $\rho_{s,RR}(L_{RR})$ is the (parallel) stiffness of a single slab of
thickness $L_{RR}$. It is related to the value of the exponent $\eta$ via
$\eta=T/(2\pi \rho_{s,RR})$. Because the probability $w(L_{RR})$ decays exponentially
with increasing $L_{RR}$, the integral is dominated by its lower bound.
To leading exponential accuracy, the parallel stiffness is thus given by
\begin{equation}
\rho_{s,\parallel} \sim \exp[-\tilde c L_c(T)] \sim \exp[- a\, (T_u-T)^{-\nu}]
\label{eq:stiff_para}
\end{equation}
with $a$ a non-universal constant. This means, it is non-zero anywhere in the
Griffiths phase and develops an exponential tail towards $T_u$ (see Fig.\ \ref{Fig:theory}).

When the twist $\theta$ is applied between the bottom ($z=0$) and the top ($z=L_\perp$) of the sample, the local
 twists between consecutive layers will vary from layer to layer. Minimizing the elastic
 free energy with respect to these local twists leads to
\begin{equation}
\rho_{s,\perp}\sim \langle1/J_{\rm eff}^\perp\rangle^{-1}
\end{equation}
where $J_{\rm eff}^\perp$
are the effective couplings between the rare regions, and $\langle \ldots \rangle$ is
the average over the sample. Because the spatial positions
of the rare regions are random, the distribution of their nearest-neighbor distances
is a Poisson distribution, $P(R) =
R_{KT} \exp(-R/R_{KT})$, where $R_{KT} \sim w(L_c)^{-1} \sim \exp[\tilde c L_c(T)]$ is the typical
separation. The effective interaction between neighboring rare regions decays
exponentially, $J_{\rm{eff}}^\perp (R) \sim \exp(-R/\xi_0)$, where $\xi_0$ is the bulk
correlation length. We thus arrive at a power-law distribution
\begin{equation}
\bar P (J_{\rm{eff}}^\perp) \sim (J_{\rm{eff}}^\perp)^{1/z -1}~.
\label{eq:J_eff-distrib}
\end{equation}
of the effective interactions.
The Griffiths dynamical exponent $z\equiv R_{KT}/\xi_0$ takes the value $\infty$ at
$T_u$, and decreases with decreasing temperature.
Using this distribution, we find that the average $\langle1/J_{\rm eff}^\perp\rangle$
diverges as long as $z>1$. This implies that the perpendicular stiffness $\rho_{s,\perp}$
vanishes in part of the Griffiths phase,
viz., between $T_u$ and a temperature $T_s$ at which $z$ reaches the value $1$.
In this temperature region, the elastic free energy density displays an anomalous dependence
on the system size, $f(\Theta) -f(0) \sim  L_\perp^{-1-z}$ corresponding
to $\rho_{s,\perp} \sim L_\perp^{1-z}$.
For $T<T_s$ the average $\langle1/J_{\rm eff}^\perp\rangle$ converges, leading to
a nonzero perpendicular stiffness. Close to $T_s$, the perpendicular stiffness
is expected to behave as $\rho_{s,\perp} \sim T_s-T$ (see Fig.\ \ref{Fig:theory}).

Other quantities can be found along the same lines \cite{MGNTV10}. For example,
the spontaneous magnetization is nonzero for all $T< T_u$ and shows
a double-exponential tail towards the nonmagnetic phase. Close to $T_u$, it takes the asymptotic form
\begin{equation}
\ln(m) \sim -\exp[a (T_u-T)^{-\nu}]~, \qquad (T\to T_u-)~.
\label{eq:m-T}
\end{equation}
If a magnetic field $H$ is applied at temperatures $T\lesssim T_u$, the magnetization-field
curve  takes the unusual form
\begin{equation}
\ln(m) \sim - \sqrt{|\ln(H)| (T_u-T)^{-\nu}}~,    \qquad (H\to 0)~, \label{eq:m-H}
\end{equation}
for magnetizations larger than the double-exponentially small spontaneous magnetization
(\ref{eq:m-T}).

For the comparison between theory and Monte-Carlo simulations, finite-size effects are an important
issue. As an example, we discuss the dependence of the magnetic susceptibility in the Griffiths
phase on the in-plane system size $L_\parallel$.  When $L_\parallel$ is finite, the susceptibility
of a single slab in the quasi-long-range ordered phase is no longer infinite.
Its $L_\parallel$-dependence can be obtained
from integrating the correlation function (\ref{eq:KT-power}) to the upper cutoff $L_\parallel$
which results in $\chi_{RR}(L_\parallel) \sim L_\parallel^{2-\eta}$. Summing this over all
rare regions yields the total susceptibility (per unit volume) as
\begin{equation}
\chi \sim L_\parallel^2 \exp\{-[\tilde c L_c(T) \ln(L_\parallel)]^{1/2}\} ~.
\label{eq:chi-L}
\end{equation}

\section{Monte-Carlo simulations}
\label{sec:simulations}

In this section, we report the results of Monte-Carlo simulations of the randomly
layered XY model (\ref{Eq:Hamiltonian}) by means of the highly efficient Wolff cluster
algorithm \cite{Wolff89}. To capture the physics of the rare regions,
we have simulated large system sizes of up to $L_\perp=800$ and $L_\parallel=100$.
In the binary distribution ({\ref{BinaryDist}}) for the
in-plane interactions $J_z^\parallel$, we have chosen the values
$J_u=1$ and $J_l=0.25$.
All the simulations have been performed for an impurity concentration $c=0.8$.
With these parameter choices, the Griffiths phase ranges from
$T_l\approx 0.97$ to $T_u\approx 2.20$.  All data are averages
over a large number (from 100 to 300) of disorder realizations. For each realization
we have used 100 Monte-Carlo (Wolff) sweeps for equilibration and
another 100 sweeps for measurements.

To test the anomalous elastic properties predicted in part of the Griffiths phase,
we have computed the parallel and perpendicular spin-wave stiffnesses.
Finding the stiffnesses by actually performing simulation runs with twisted boundary
conditions is not very efficient. However, the stiffnesses can be rewritten in terms
of expectation values calculated in a conventional run with periodic boundary conditions.
In the case of the perpendicular stiffness, the resulting formula
\cite{TeitelJayaprakash83} (see also \cite{HrahshehBarghathiVojta11})
reads
\begin{eqnarray}
 \rho_{s,\perp}&=&\frac 1 N \sum_{\langle\mathbf r,\mathbf r'\rangle}J_{\mathbf r, \mathbf r'}
                 \langle \mathbf S_{\mathbf r}\cdot\mathbf S_{\mathbf r'}\rangle (z-z')^2 \\
&&-\frac{1}{NT}\left\langle\left(\sum_{\langle\mathbf r,\mathbf r'\rangle}J_{\mathbf r, \mathbf r'} \, \hat {\mathbf k}\cdot (\mathbf S_{\mathbf r}\times \mathbf S_{\mathbf r'})
 (z-z')\right)^2\right\rangle~,\nonumber
\end{eqnarray}
where $N=L_\perp L_\parallel^2$ is the total number of sites, and $\hat {\mathbf k}$ is a unit vector
perpendicular to the plane of the XY spins. For the calculation of $\rho_{s,\parallel}$, the term
$(z-z')$ needs to be replaced by $(x-x')$.

\begin{figure}
\includegraphics[width=8.cm]{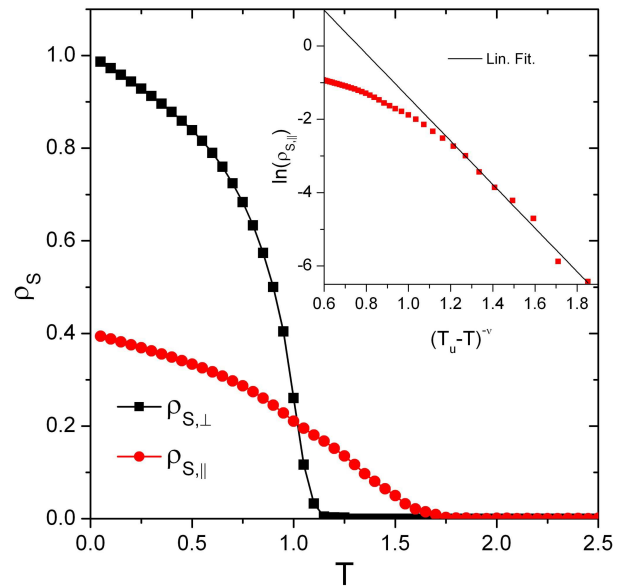}
 \caption{(Color online) Perpendicular and parallel spin-wave stiffnesses,
  $\rho_{s,\perp}$ and $\rho_{s,\parallel}$, as functions of temperature $T$
 for a system with sizes $L_\perp=800$ and $L_\parallel=100$. The data are averaged over 100 disorder configurations.
 The inset shows that $\rho_{s,\parallel}$ follows the prediction (\ref{eq:stiff_para})
 for $T\to T_u$.}
 \label{fig:stiffnesses}
 \end{figure}

Using this formula, we have calculated the parallel and perpendicular stiffnesses
of a system of sizes $L_\perp=800$ and $L_\parallel=100$. The results are shown in
Fig.\ \ref{fig:stiffnesses}.
The two stiffnesses clearly behave differently. The perpendicular stiffness $\rho_{s,\perp}$
vanishes at a temperature $T_s \approx 1.15$ while the parallel stiffness $\rho_{s,\parallel}$
remains nonzero to significantly higher temperatures and develops a tail towards
$T_u$. In agreement with the theoretical predictions, we thus find an intermediate
anomalously elastic (sliding) phase in which $\rho_{s,\perp}=0$ but $\rho_{s,\parallel} \ne 0$.
To further test the theory, we plot $\ln(\rho_{s,\parallel})$ vs. $(T_u-T)^{-\nu}$ in the inset
of Fig.\ \ref{fig:stiffnesses}. According to (\ref{eq:stiff_para}), the data sufficiently
close to $T_u$ should fall onto a straight line. As the inset shows, our results follow the
prediction over more than 1.5 orders of magnitude in $\rho_{s,\parallel}$ (down to the resolution
limit set by the Monte-Carlo noise).

\begin{figure}
\includegraphics[width=8.5cm]{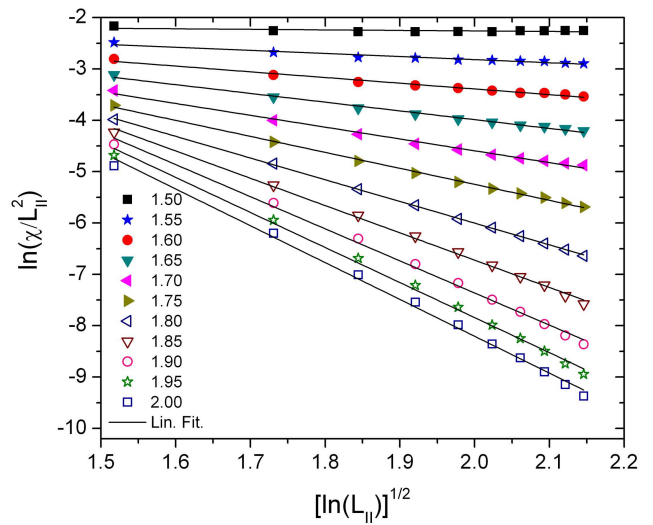}
\caption{(Color online) Susceptibility $\chi$ as a function of in-plane system size $L_\parallel$
for several temperatures in the Griffiths region. The perpendicular size is $L_\perp=800$;
the data are averages over 300 disorder configurations. The solid lines are fits
to the theoretical prediction (\ref{eq:chi-L}).}
\label{fig:chiL}
\end{figure}

In addition to the spin-wave stiffnesses, we have also analyzed the finite-size
behavior of the magnetic susceptibility.
Figure \ref{fig:chiL} shows the susceptibility $\chi$ as a function of $L_\parallel$
for several temperatures in the Griffiths region.
We have used system sizes $L_\perp=800$ and $L_\parallel=10$ to $100$
because the condition $L_\perp \gg L_\parallel$ (such that $L_\perp$ is effectively infinite)
needs to be fulfilled
when studying the dependence of $\chi$ on $L_\parallel$.

To compare the simulations to the theory, we plot the Monte-Carlo data
as $\ln(\chi/L_\parallel^2)$ vs. $[\ln(L_\parallel)]^{1/2}$. In such a plot,
the functional form (\ref{eq:chi-L}) yields a straight line.
Figure \ref{fig:chiL} shows that our data are in good agreement
with the theoretical prediction over a wide temperature range. Some deviations
appear for temperatures close to $T_u$ and large $L_\parallel$. They are
likely caused by the fact that the perpendicular size $L_\perp$ is not truly
infinite in our simulations. Thus, a typical sample will not contain any of the very
thick rare regions that dominate $\chi$ at temperatures close to $T_u$.
From fits of the data to (\ref{eq:chi-L}), one can obtain estimates of the cutoff length
scale $L_c(T)$. Its temperature dependence does not agree very well with
the prediction $L_c \sim (T_u-T)^{-\nu}$, probably because the theory holds
asymptotically close to $T_u$ while the simulations for such temperatures
suffer from the finite system size $L_\perp$, as discussed above.

\section{Conclusions}
\label{sec:conclusions}

In summary, we have performed large-scale Monte-Carlo simulations of a classical
three-dimensional XY model with layered randomness. Our results provide support for
the recently predicted unconventional phase transition scenario \cite{MGNTV10,PekkerRefaelDemler10}.
In particular, we have found evidence for the existence of an
anomalously elastic (sliding) intermediate phase between $T_u$ and $T_s$.
In this phase, the stiffness parallel
to the layered randomness is nonzero while the perpendicular stiffness vanishes.
We have also confirmed the unusual finite-size scaling behavior of
the magnetic susceptibility.

It is interesting to compare the present results for a randomly layered XY model
with corresponding results for Ising and Heisenberg spins. In a randomly layered
Ising model, each rare region (slab) corresponds to a two-dimensional Ising model;
it can thus undergo a transition to a long-range ordered state independently
of the bulk system. The global phase transition is therefore smeared
\cite{Vojta03b,SknepnekVojta04}. (In the Ising case, the tail of the magnetization
towards $T_u$ decays as a single exponential, i.e., $m$ is significantly larger than
in the double exponential tail (\ref{eq:m-T})). In contrast, for Heisenberg symmetry, the rare regions
correspond to two-dimensional Heisenberg models which \emph{cannot} develop long-range
order independently. As a result, the global phase transition is sharp. However,
the layered randomness leads to exotic infinite-randomness critical behavior
\cite{MohanNarayananVojta10,HrahshehBarghathiVojta11}.
In our case of the randomly layered XY model, the rare regions can undergo a phase transition
independently from each other, but only to a quasi-long-range ordered state rather than
to true long-range order. In terms of the classification
\cite{VojtaSchmalian05,Vojta06} of phase transitions in disordered systems,
the randomly layered XY model is thus features a hybrid between a smeared and
a sharp transition.

Turning to experiment, our results are applicable to a variety of systems.
Even though our theory is formulated in the language of the planar (XY) ferromagnet, it
holds for all thermal phase transitions with O(2) or U(1) order parameters, if it is
expressed in terms of the appropriate variables. For randomly layered superconductors
and superfluids, the magnetization should be exchanged for the Cooper pair amplitude
or the condensate wave function. Analogously, the spin-wave stiffness should be substituted
by the superfluid density. It is worth noting that recent large-scale
quantum Monte-Carlo simulations of the Bose-Hubbard Hamiltonian on randomly stacked
layers \cite{Laflorencie12} also confirmed the existence of an anomalously elastic (sliding)
phase between the normal fluid and the superfluid in this system.

Experiments in ultracold atomic gases have already demonstrated the Kosterlitz-Thouless transition
in stacks of quasi two-dimensional layers created by a strong one-dimensional optical lattice
\cite{HKCBD06}. Moreover, disordered optical lattices have been used to study
Anderson localization of matter waves \cite{Billyetal08,Roatietal08}. Our results apply to
large irregular stacks of quasi two-dimensional layers created by a one-dimensional
disordered optical lattice of a strength that still allows some weak coupling between the layers.

Recently, experiments on several layered perovskite superconductors
\cite{LHGTT07,DSBBK11,Wenetal12} found unexpected anisotropies
of the superconducting properties that imply an apparent decoupling of the
superconducting layers. Our anomalously elastic phase may explain these
observations provided that there is sufficient c-axis disorder is the samples.
Our theory could also be tested by manufacturing layered nanostructures of different
magnetic or superconducting materials. Magnetic multilayers with systematic variations of the critical temperature
from layer to layer have already been produced \cite{MPHW09}. The system we have studied
can be realized as a random version of such a structure using an easy-plane magnetic
material.

\section*{Acknowledgements}

We acknowledge helpful discussions with N. Laflorencie, R. Narayanan, D. Pekker, and G. Refael.
This work has been supported in part by the NSF under grant no. DMR-0906566.

\bibliographystyle{apsrev4-1}
\bibliography{rareregions}
\end{document}